# Shape and motion of drops in the inertial regime

Vijaya Senthil Kumar K[1] and Baburaj A. Puthenvettil[1*]

[1]Fluid Mechanics Laboratory, Department of Applied Mechanics, Indian Institute of Technology Madras, Chennai, India.

*E-mail of the presenting author: apbraj@iitm.ac.in

**Abstract** In this paper, we report experimental results on the shape and motion of a mercury droplet, placed in a horizontally rotating cylinder in the rpm range 8-93, so that the Reynolds number of the drop is 2500<Re<26000 and its capillary number is 0.0002<Ca<0.0023. When contact angle variations can be neglected at low speeds (Re<8150), the velocity of the drop is much lower than that predicted by the Ho. Young Kim's [6] relation. This observed discrepancy is overcome by modifying Kim's relation by substituting the dissipation estimated from a boundary layer near the solid surface instead of bulk dissipation. Based on the changes at the rear side of the mercury droplet, there are three distinct regimes identified with varying speeds of rotation (i) oval or rounded regime (ii) corner regime and (iii) cusping regime. The oval to corner transition happens at a finite receding contact angle of $95^0$. The ratio of critical contact angle ($\Theta_c$) at which the transition occurs to the static receding contact angle ($\theta_s$) was found to be 0.657. The de Gennes model [4], extended to high contact angle by substituting the dissipation for wedge flow, predicts a critical contact angle ratio ($\theta_c/\theta_s$) that is in close agreement with the experimental value. At higher Re, the dynamic contact angle variation with velocity was compared with Cox-Voinov model [3]. Though the trend of the variation of data is approximately represented by the model, the fit coefficient according to experimental data is very high when compared to theoretical value.

## 1. Introduction

The steady motion of a drop on an inclined surface occurs at a velocity that balances the gravitational driving forces with the viscous resistance and the resistance at the contact line [6]. However, a complete understanding of the phenomena and the regimes of drop motion is still elusive due to incomplete knowledge of the motion of the three phase contact line. In this paper, we investigate a hitherto unexplored regime of drop motion at large Reynolds number Reynolds number (Re) and contact angles, but at very small capillary numbers (Ca) by using mercury over glass surface. Here, Re=Ud/ν, where U is the drop velocity, d is the diameter of the drop and ν is the kinematic viscosity and Ca = μU/σ, where μ is the dynamic viscosity and σ is the surface tension of the fluid. An understanding of such a regime of motion of drops on surfaces is of key importance in micro fluidics, surface coatings, oil recovery from bed rocks, efficient deposition of pesticides on plant leaves, ink jet printing, industrial condensers employing drop wise condensation etc., Dussan and Davis [5], by visualising the motion of a surface marker on a drop of honey, were the first to clarify that the kinematics of drop motion on inclined surfaces is predominantly rolling. At low velocities of the drops, by neglecting the contact angle variation with velocity, a global energy balance between the gravitational potential energy $\Phi_g = \rho Vg\sin\alpha$ and total dissipation inside the drop $\Phi_t$ could be written down to obtain various expressions for the velocity of the drop U for various predominant dissipation mechanisms [6]. Here, ρ is the density of the fluid, V is the volume of the drop and α is the inclination angle and g is the acceleration due to gravity. The expression for large drops when $\Phi_t = \Phi_w+\Phi_b+\Phi_{cl}$, where $\Phi_w$ is the wedge dissipation along the circumference of the droplet, $\Phi_b$ is the bulk dissipation and $\Phi_{cl}$ is the contact line dissipation, was given by [6] as,

$$U \approx \frac{\rho gV \sin\alpha - \sigma w(\cos\theta_r - \cos\theta_a)}{\mu(V_b/h^2 + Lc(\theta_s)\ln(R_b/\lambda))} \quad (1)$$

Here, σ is the surface tension, $R_b$ is the radius of the base of the droplet h is the height of the droplet and $\theta_a$ and $\theta_r$ are the static advancing and receding contact angles respectively. $V_b$ is the volume of the base of the droplet which is given by $\pi R_b h$, L is the circumference of the droplet which is approximately equal to $2\pi R_b$, w is the width of the droplet and λ is the microscopic cut-off length scale where the continuum approximation fails, typically of the order of $10^{-9}$ m. $c(\theta_s)$ is a function of contact angle. They verified their expressions for the case of small drops (h<$\sqrt{\frac{\sigma}{\rho g}}$) with experiments using viscous fluids. Further studies by Podgorski et al. [9] on the shape transition of silicon oil drops showed that the drops develop a corner and then a cusp at the rear at two critical speeds. Le Grand et al. [7] improved upon the results of [9] by studying silicon oil drops over a wide range of viscosities (10cP-1000cP) and found that, none of the existing contact angle models were universally consistent with their results; the Cox-Voinov law,

$$\theta_d^3 - \theta_{s,r}^3 = \pm 9Ca\ln(R/\lambda) \quad (2)$$

where $\theta_d$ and $\theta_{s,r}$ are the dynamic and static receding contact angles and R is the macroscopic length scale, was found to be the most consistent with their experimental data. One important result of their work was that the corner transition happened at a non-zero receding contact angle of 23° contrary to the assumptions of Blake and Ruschak [1] and Podgorski et al. [9]. However, the Cox-Voinov law did not predict the corner transition at a nonzero receding contact angle while the de Gennes [4] model,

$$\theta_d(\theta_d^2 - \theta_{s,r}^2) = \pm 6Ca\ln(R/\lambda) \quad (3)$$

did so. All of these studies are conducted with viscous liquids that partially wet the solid surface so that the drops move slowly (<1cm/s) with a Re ~ 1.

Very little is known quantitatively about the motion of partially non-wetting of drops at high Reynolds number. This regime in addition to being of great applicational value in the fields like high throughput micro fluidics and high speed coatings is of interest in understanding the inertial effects in drop motion. High contact angles on smooth surfaces occur due to high surface tension, resulting in low Ca. So motion of partially non-wetting drops exhibit a regime of low Ca and high Re, implying a predominance of inertial effects. We achieve such a regime by using mercury drop (static contact angle, $\theta_s$ ~140°) on a glass surface. The Reynolds number range achieved is 2500<Re<26000 while capillary numbers are in the range of 0.0002<Ca<0.0023. Since the drops move at speeds of order 1m/s we freeze the drops relative to the observer by placing the drop inside a sealed glass cylinder rotating about a horizontal axis.

The paper is organised as follows. The experimental setup is discussed in section 2. The velocity of the droplet and its shape transition is discussed in section 3 and section 4 respectively. The dynamic contact angle variation under negligible contact angle variation with velocity is discussed in section 5 and conclusion drawn from the discussion is given in section 6.

## 2. Experimental setup

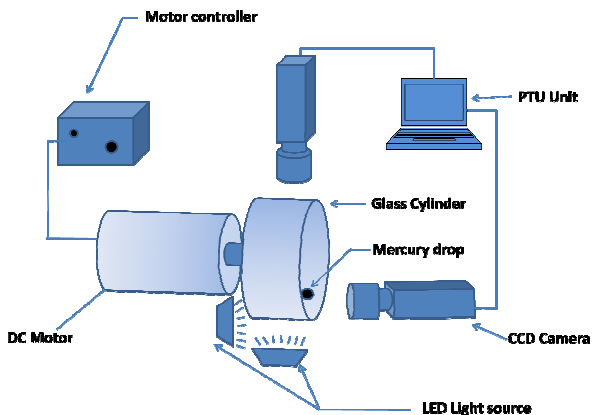

Figure 1: Schematic of the experimental setup

The setup consists of a horizontally rotating glass cylinder of radius 7.5 centimetres which is connected to the shaft of a variable speed electric motor as shown in figure 1. The mercury drop is placed inside the cylinder and sealed. Pure mercury droplets of diameter 4 millimetres having a mass of 0.022 g and a smooth quartz glass cylinder were used in the experiment. The properties and dimensions of the mercury droplet are listed in table 1. For each rotation speed and volume, the drop moves along the cylinder and becomes stationary when a balance among viscous, gravitational and interfacial forces is achieved. When the drop size is small compared to the radius of cylinder, the phenomena will now be analogous to steady motion of a drop on an inclined plate. The analogous inclination angles can be calculated from the side views of the image captured by CCD camera by fitting a tangent at the drop's base and measuring the angle between the horizontal line and the tangent. The contact angles are measured from the side views of the image of the drop by fitting two tangent lines at the contact line. The error in contact angle for mercury drop is found to be 3 degrees from repeated measurements.

| μ (Pa.s) | σ (Nm) | ρ kgm$^{-3}$ | $R_b$ mm | w mm | $\Theta_a$ deg | $\Theta_r$ deg |
|---|---|---|---|---|---|---|
| 0.001526 | 0.48545 | 13526 | 1.75 | 4 | 150 | 144.5 |

Table 1: Properties and dimensions of mercury droplet

The cylinder speed at which the drop comes to the final position of rest gives us the relative velocity of the drop U with respect to the moving surface. The rotational speed of the cylinder was measured using a digital tachometer. The contact angle is easily measured as the drop would be stationary with respect to the observer. The cameras were operated at 25 frames per second and the resolution is of the order of 10$^{-4}$m per pixel. Two CCD cameras connected with programmable timing unit (La Vision) were used for visualization of the top and side view of the drop. Two LED lights were used as a source of back lighting one at the back of the cylinder and another below the cylinder opposite to the CCD cameras.

## 3. Velocity of the droplet with negligible contact angle variation

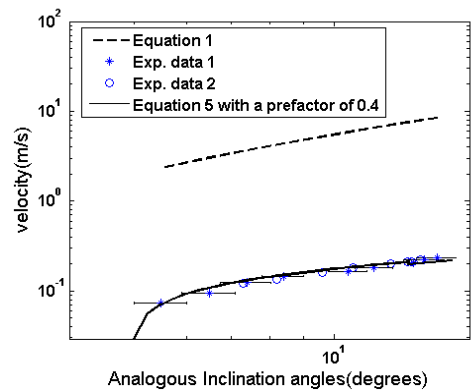

Figure 2: Comparison between experimental values and theoretical relations

Figure 2 shows the measured velocities of the mercury drop as a function of the corresponding inclination angle α for the case of an inclined flat plate. The errors in the calculated inclination angles shown in figure 2 are estimated by repeated measurements of inclination angle for each rotation rate. The theoretical prediction by equation (1) using the values listed in table 1 is also shown in the figure 2. The velocity of the droplet is lower than that predicted by equation (1) by one order of magnitude. We expect this discrepancy to be due to the estimate of bulk viscous dissipation used in the equation, which is valid only at low Re. Inertial effects cannot be neglected in the case of high dense mercury droplets whose motion in the present study result in a Re range of 2500<Re<26000. In this inertial regime, we expect the viscous dissipation to be concentrated in a thin boundary layer near the solid surface. The boundary layer thickness inside the droplet scales as [8],

$$\delta \approx (\mu R / \rho U)^{1/2} \quad (4)$$

Using $V_{bl} = \pi R_b \delta$, instead of $V_b$ in equation (1), with $\delta$ given by equation (4), the modified equation for the sliding velocity would be,

$$\pi R_b (\rho \mu R_b)^{0.5} U^{2.5} + \mu L c(\theta) U \ln(R_b / \lambda) - (\rho V g \sin\alpha - \sigma w(\cos\theta_r - \cos\theta_a)) \approx 0 \quad (5)$$

Equation (5) is solved numerically to obtain the velocity of the drop shown in figure 2. The theoretical prediction of the equation (5) with a prefactor of 0.4 represents the experimental values better than that by equation (1).

## 4. Shape & wetting transition

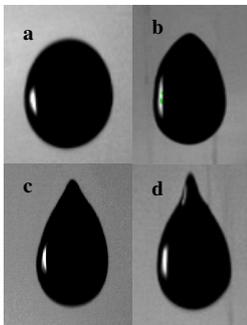

Figure 3: Drop shape at various Ca (a) Oval or rounded at the top at Ca 0.0013 (b) corner at Ca 0.0016 and (c) (d) cusping at 0.0025 and .003 respectively.

The mercury drop undergoes shape transition when the capillary number is increased as shown in the figure 3. Based on the shape changes at the receding side, three distinct regimes can be observed at varying speeds of rotation of the cylinder, (i) oval or rounded regime (figure 3a) (ii) corner regime (figure 3b) and (iii) cusping regime (figure 3c). Beyond cusping regime, droplets start emitting from the rear due to instabilities similar to Plateau-Rayleigh instability of a jet. The transition from a rounded receding contact line to a corner happens at Ca = 0.0015, and cusping starts at Ca = 0.0021. The corner does not have an exact conical structure, but resembles a cone with a rounded tip as shown in the figure 3(b). This formation of corner at the rear of the droplet, inferred as the wetting transition, happens at a finite receding contact angle $\theta_c$ as can be seen from figure 4(a). The ratio $\theta_c/\theta_s$ is 0.657 for mercury drop on glass, higher than 0.577 predicted by the de Gennes model [4]. The idea behind the de Gennes model is to equate the viscous dissipation $\Phi_{vis}$ near the contact line to the dissipation due to the non equilibrium force acting on the contact line $\Phi_{cl}$, which is moving with a velocity $U_c$. It is known that the

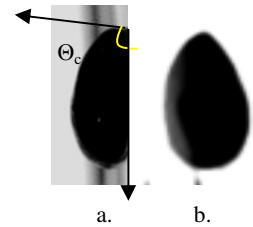

a.   b.

Figure 4: Corner formation at the finite receding contact angle of 95°

contact angle hysteresis increases the critical value at which the corner appears [10]. However this cannot be the case here as mercury on glass has low hysteresis (6°). We hence, expect this discrepancy to be because the de Gennes model is limited to a wedge angle less than one radian, which is clearly not the case with mercury droplets. The de Gennes model can be extended to high contact angle liquids by substituting the viscous dissipation $\Phi_{vis}$ from a wedge with a dynamic contact angle $\theta_d$ instead of that from a Poiseuille flow in the original model [4]. The resulting relation between contact line velocity $U_c$ and $\theta_d$ is given as,

$$U_c = \frac{\sigma(\cos\theta_s - \cos\theta_d)(2\theta_d - \cos\theta_d \sin\theta_d)}{4\mu \ln(R/\lambda)\sin^2\theta_d} \quad (6)$$

The above equation predicts a maximum velocity at a finite critical contact angle $\theta_c$, between $\theta_s$ and zero, as shown in figure 5. This critical contact angle can be found by differentiating the above equation and equating it to zero. The critical contact angle ratio ($\theta_c/\theta_s$) obtained in such a way from this relation for mercury is 0.6354 which is close to 0.657 observed from experiments.

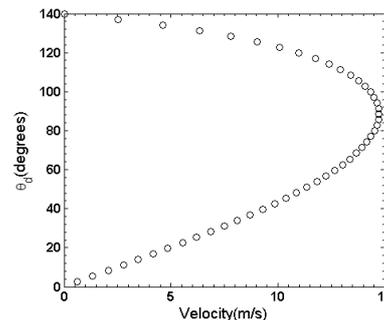

Figure 5: Dynamic contact angle vs velocity according to equation 6.

## 5. Dynamic contact angle variation

The advancing contact angle increases and the receding contact angle decreases with increase in velocity respectively as shown in figure 7. The algebraic capillary number $\overline{Ca}$ in figure 6 is defined as $\mu U/\sigma$, where U is positive for advancing contact angle and negative for receding case. The receding contact angle becomes almost constant in the corner and the cusping regime, as can be seen from figure 6. The Cox-Voinov [3] relation between the contact angle and the capillary number, for contact angle of values less than 135°, is given by equation 2.

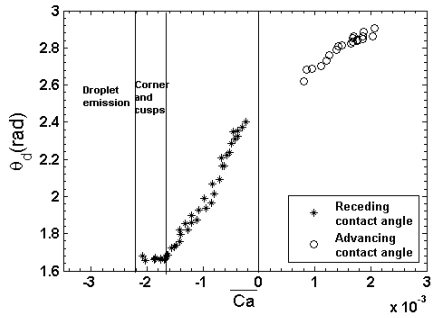

Figure 6: Dynamic contact angle Vs Algebraic Capillary number $\overline{Ca}$.

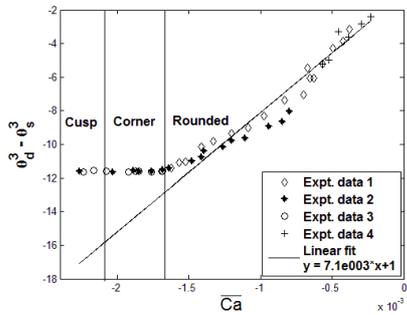

Figure 7: Comparison of experimental data with Cox-Voinov relation

Figure 7 shows the comparison between the experimental receding contact angles and the theoretical Cox-Voinov relation. Though the trend is cubic up to the corner transition point, the experimental value of the constant $9\ln(R/\lambda)$ is 710 as against the expected theoretical value of 13 [4]. Contrary to the present experimental observations, according to Cox-Voinov relation, the contact angle change should be very minimum for the range of capillary numbers of the experiment. Beyond corner transition, the theoretical curve does not clearly represent the experiment well.

Cox-Voinov is a slip model for contact line dynamics valid only when the local Re near the contact line is less than 1. The model assumes that the actual contact angle at the contact line is $\theta_s$, the measured dynamic contact angle being considered as apparent due to the finite resolution of all measurements. This assumption necessitates the presence of stagnant fluid zone of length $\lambda$ near the contact line across which the fluid velocity changes from that of the solid to that of the liquid-gas interface [11]. The present results show that, even if we consider a very small length scale near the contact line so that the local Re<1 to apply the Cox-Voinov model, the slip length $\lambda$ will go beyond the continuum length scales. The present study hence brings out the limitation of the slip models of contact line dynamics.

## 6. Conclusions

In this study, we measured the velocities and the contact angle variation of mercury drops for a range of Re and Ca by using a rotating cylinder arrangement. It was found that for negligible contact angle variation from static values (Re<8150), the velocity of the drop does not obey Ho.Young Kim's [6] relation based on Stokes flow analysis. We proposed a new relation by including the boundary layer dissipation in the energy balance, which predicted the velocities reasonably. At larger Re (8150<Re<26000) and Ca (0.0005<Ca<0.002), the dynamic contact angle varies appreciably from the static contact angle with increase in velocity; this variation is not captured by the Cox-Voinov [3] relation (equation 2). This implies the unavoidable need for the inclusion of contact line physics in the analysis of problems involving moving contact line. The drop exhibits shape transition with increasing velocities. The oval to corner transition at the rear of the droplet happens at a finite receding contact angle $\theta_c$ of 95°. The critical contact ratio $\theta_c/\theta_s$ for the mercury drop is 0.657. We extend the de Gennes model to high contact angle cases to show that $\theta_c/\theta_s$ predicted by the new model is close to the experimental ratio. This observed first order wetting transition at high contact angle cases implies the dominant nature of viscous dissipation near the contact line, contradicting previous conclusion made in [2].